\documentclass{aastex63}
\usepackage{color}

\usepackage{graphicx}
\usepackage{subfigure}
\usepackage{epsfig}
\usepackage{rotating}
\usepackage{floatrow}
\DeclareFloatFont{scriptsize}{\scriptsize}
\newcommand{\Erot}{E_{\rm rot}}
\newcommand{\Edotrot}{\dot{E}_{\rm rot}}
\newcommand{\Edotem}{\dot{E}_{\rm em}}
\newcommand{\Edotgw}{\dot{E}_{\rm gw}}
\newcommand{\Edotgwr}{\dot{E}_{\rm gw,r}}

\newcommand{\gammar}{\gamma_{\rm r}}

\newcommand{\tgwr}{\tau_{\rm gw,r}}

\shortauthors{Jie Lin et al.}

\begin{document}

\title{GRB~111209A/SN~2011kl: Collapse of a supramassive magnetar with r-mode oscillation and fall-back accretion onto a newborn black hole}

\correspondingauthor{Rui-Jing Lu}
\email{luruijing@gxu.edu.cn}

\author{Jie Lin}
\affiliation{Guangxi Key Laboratory for Relativistic Astrophysics, School of Physical Science and Technology, Guangxi University, Nanning 530004, People's Republic of China}

\author{Rui-Jing Lu}
\affiliation{Guangxi Key Laboratory for Relativistic Astrophysics, School of Physical Science and Technology, Guangxi University, Nanning 530004, People's Republic of China}

\author{Da-Bin Lin}
\affiliation{Guangxi Key Laboratory for Relativistic Astrophysics, School of Physical Science and Technology, Guangxi University, Nanning 530004, People's Republic of China}

\author{Xiang-Gao Wang}
\affiliation{Guangxi Key Laboratory for Relativistic Astrophysics, School of Physical Science and Technology, Guangxi University, Nanning 530004, People's Republic of China}










\begin{abstract}
Ultra-long-duration gamma-ray burst GRB~111209A was found to be associated with a very luminous supernovae (SNe) SN~2011kl. The physics of GRB~111209A/SN~2011kl has been extensively studied in the literures, but does not settle down yet. By investigating in detail the characteristics of the X-ray light curve of GRB~111209A, coupled with the temporal and spectral features observed in SN~2011kl, we argue that a short-living supramassive magnetar can be responsible for the initial shallow X-ray emission. Then the electromagnetic extraction of spin energy from a black hole results in the steeply declining X-ray flux when the magnetar collapses into a black hole (BH). A fraction of the envelope materials falls back and activates the accretion onto the newborn BH, which produces the X-ray rebrightening bump at late times. During this process, a centrifugally driven baryon-rich quasi-isotropic Blandford \& Payne outflow from the revived accretion disk deposits its kinetic energy on the SN ejecta, which powers luminous SN 2011kl. Finally, we place a limitation on the magnetar's physical parameters based on the observations. 
\end{abstract}



\section{Introduction}
	
Gamma-Ray Bursts (GRBs) are the most luminous explosions in the universe since the big bang, and now are commonly divided into short GRBs and long GRBs (LGRBs), with a separation around $T_{\rm 90} \approx 2 s$ (where $T_{\rm 90}$ is the time window containing 90\% of the energy released, \citealt{1993ApJ...413L.101K}). Short GRBs have been widely thought to be produced by mergers of compact binary systems composed of two neutron stars (NSs) or a NS and a black hole (BH) \citep{1989Natur.340..126E,1991AcA....41..257P,1992ApJ...395L..83N}. The directly evident of compact binary mergers being the progenitors of short GRBs is that, recently, the coincident detection of a gravitational wave event GW170817 and its electromagnetic counterparts, i.e., a short GRB 170817A and a kilonova \citep{1998ApJ...507L..59L,2017ApJ...848L..13A,2017ApJ...848L..14G,2017Natur.551...64A,2017Natur.551...67P,2017Natur.551...75S,2017Natur.551...80K,2017Sci...358.1559K}. Whereas LGRBs are commonly believed to originate from the death of massive stars \citep{1999ApJ...524..262M,macfadyen01}, and the solid confirmation came with the secure spectroscopic association of a broad-lined Type Ic SN (2003bh) with GRB 030329 \citep{2003Natur.423..847H,2003ApJ...591L..17S}. Alternatively, some authors also argued that massive stars in compact binaries are a likely channel for some fraction of LGRBs (e.g., \citealt{2006MNRAS.372.1351L,2012MNRAS.425..470C,2017AdAst2017E...5C}). 

In the last several years, a handful of GRBs lasting longer than $10^3$ s have been discovered and are now called ultra-long GRBs (ULGRBs). ULGRBs' durations are statistically different from that of LGRBs \citep{2015JHEAp...7...44L,2017ApJ...839...49G}. If ULGRB originates from the core collapse of massive stars, its central engine falls into two classes: a hyperaccreting stellar-mass BH \citep{1993ApJ...405..273W} or a NS with millisecond rotation period and an ultra-strong magnetic field \citep{1992Natur.357..472U,dai12,2001ApJ...552L..35Z,2017AdAst2017E...5C,2017ApJ...839...49G}. 

But what physical mechanism and central engine are responsible for the ULGRBs is now unclear yet. GRB~111209A, one of very rare ULGRBs, was found to be associated with SN~2011kl, one of the most luminous GRB-associated supernova \citep{2015Natur.523..189G}. The abnormal luminous SN~2011kl is intermediate between canonical GRB-associated supernova and superluminous supernova (SLSNe; \citealt{2011Natur.474..487Q,galyam12,2016ApJ...826..141G,2017ApJ...850..148W}), and resembles SLSNe than any classical known GRB-SN, such as the slope of its's continuum resembles those of SLSNe but extends further down into the rest-frame ultraviolet implying a low metal abundance \citep{2015Natur.523..189G,2016MNRAS.458.3455M,2019A&A...624A.143K}. Such a fact may provide a striking clue to the nature of ULGRBs associated with SLSNe. 

\citet{2015Natur.523..189G} argued that the decay of $^{56}$Ni can not be responsible for powering the luminosity of SN~2011kl since a large $^{56}$Ni mass is required, which is inconsistent with the low metal abundance. It suggests that an additional energy input is required. Furtherly, they found that the bolometric light curve of SN~2011kl can be reproduced by a model where extra energy is injected by a magnetar, which has also been proposed as the energy source of SLSNe. This means that the central engine of GRB 111209A may be a magnetar. Alternatively, other models, e.g., $^{56}$Ni cascade decay model, magnetar spin-down model, ejecta-circumstellar medium interaction model, and jet-ejecta interaction model, are also proposed to interpreted the light curve of a SLSNe (detailed information refered to \citealt{wang19}).

However, when investigating the X-ray afterglow observations of GRB 111209A in detail, one could found that the X-ray light curve  exhibits a sharply decay at the end of the shallow decay phase, and next follows a small bump (the same feature also seen in it's corresponding optical light curve \citealt{2017ApJ...839...49G}). By analyzing the time-resolved spectra from the XMM-Newton data within the small bump phase, \citet{2013ApJ...779...66S} revealed that a harden power-law component during this phase is need to accommodate the data, suggesting an additional energy input to explain this observation.
	
A steep decay phase at the end of shallow decay, followed by a X-ray bump, has been usually thought to the collapse of a supermassive magnetar into a BH when it spins down \citep{2007ApJ...665..599T,2010MNRAS.409..531R,2013MNRAS.430.1061R,2014ApJ...780L..21Z,2017ApJ...849..119C}. Therefore, the physical picture, that the central engine of GRB 111209A should initially be a magnetar and then the magnetar collapses into a BH after it spins down, should prefer accommodating these observations. If so, however, another question arises, how such a short-lived magnetar could be responsible for the energy source powering the luminous SN 2011 as pointed out in \citet{2015Natur.523..189G}?
	
\citet{2016ApJ...826..141G} proposed a collapsar model with fallback accretion to explain the energy source of the luminous SN~2011kl, in which the progenitor star has a core-envelope structure with the bulk of the mass in the core part collapsing into a rapidly spinning BH
and the rest mass forming a accretion disk around the BH. The prompt emission of GRB 111209A could be powered by the Blandford \& Znajek (\citealt{blandford77}, hereafter BZ) mechanism. The energy and angular momentum could also be extracted magnetically from accretions disks, by centrifugally launching a baryon-rich wide wind/outflow through the Blandford \& Payne (\citealt{blandford82}, hereafter BP) mechanism.
This quasi-isotropic outflow would further deposit its kinetic energy on the SN ejecta, which produces the luminous SN 2011. Similarly, in a merger model, such a BP-driven outflow from a newborn BH's disk can heat up, push the ejecta, and produce a merger-nova as bright as a supernova \citep{ma2018}.
	
Inspired by \citet{2016ApJ...826..141G}, together with the comprehensive
observations of GRB 111209A/SN~2011kl, we propose another new interpretation of both the X-ray light curve of GRB 111209A and the energy source of SN~2011kl in this paper. The physical picture is as follows: the initially shallow decay can be explained by that the r-mode gravitational-waves emission dominated the spin-down evolution of the magnetar as seen in short GRB 090510 \citep{2019ApJ...871..160L}, then the following abruptly decay phase results from the extraction of the black hole rotational energy through the BZ mechanism when the magnetar loses its angular momentum via r-mode gravitational-waves radiation and then spins down until the centrifugal force is insufficient to support the mass to collapse into a BH \citep{2014PhRvD..89d7302L}. Finally, the small X-ray bump at late time is likely due to fallback accretion into the newborn BH, as seen in other GRBs, such as GRB 121027A and GRB 070110 \citep{2013ApJ...767L..36W,2017ApJ...849..119C}. Meanwhile, the BP outflow from the newborn BH's disk would further deposit energy into the SN ejecta, which powers the luminous SN~2011kl.

This paper is organized as follows. We firstly summarize the observations of GRB~111209A/SN~2011kl in Section 2.
In Section 3, we describe the models applying to explain the observations of GRB~111209A/SN~2011kl.
In section 4, we briefly summarize our results and discuss the corresponding implications.
Throughout the paper, the standard notation $Q_{x} = Q/10^{x}$ with Q being a generic quantity in cgs units and a concordance cosmology with parameters $H_{0}$=71~km~s$^{-1}$~Mpc$^{-1}$, $\Omega_{\lambda}=0.73$, $\Omega_{m}=0.27$ are adopted.

\section{OBSERVATIONs OF GRB~111209A/SN~2011kl}
	
GRB~111209A was discovered by the Burst Alert Telescope (BAT) on board {\it Swift} at $T_{0}$=2011:12:09-07:12:08 UT.
The prompt emission of GRB~111209A showed an exceptionally long duration and was monitored up to $T_0+1400$ s.
GRB~111209A was also detected by the Konus-WIND.
But as shown in the ground data analysis of the Konus-Wind instrument,
this burst started about at 5400 s before $T_{0}$ and showed a weaker broad pulse of emission seen from $T_{0}$-5400 s to $T_{0}$-2600 s \citep{golenetskii11}.
According to the Konus-Wind results, it had a fluence of $(4.86 \pm 0.61) \times 10^{-4} \rm erg\cdot cm^{-2}$ corresponding to an isotropic gamma-ray energy
$E_{\gamma,\rm iso}=4\pi D_{L}^2 f_{\gamma}/(1+z) = (5.82 \pm 0.73 )\times10^{53}$ erg \citep{golenetskii11}.
	
{\it Swift}/XRT observations started at 425 s after the BAT trigger and observed a bright afterglow at a position of RA(J2000)=00$^h$ 57$^m$ 22.63$^s$ and Dec(J2000)=-46d 48$'$ 03.8$''$, with an estimated uncertainty of 0.5 arcsec \citep{hov11}.
The X-ray afterglow is also revealed by {\it XMM-Newton} in the period
between $T_{0}$+56,664 s and $T_{0}$+108,160 s.
The optical counterpart of GRB 111209A was detected by {\it Swift}/ UVOT in the optical-UV bands and other ground based instruments, such as the TAROT-La Silla \citep{klotz11} and the GROND robotic telescopes \citep{kann11}.
The early spectroscopy of the transient optical light of GRB 111209A was obtained with both Very Large Telescope (VLT)/X-shooter (2011 December 10 at 1:00 UT) and Gemini-N/GMOS-N \citep{2016ApJ...826..141G}.
A redshift of $z = 0.677$ was provided by identifying absorption lines and emission lines from the host galaxy  \citep{vreeswijk11}.
Observationally, the X-ray afterglow of GRB 111209A can be divided into mainly four phases: (I) an initial shallow decay until $\sim$7.5 ks, with three brigh flares superimposing on the underlying smooth component; (II) a sharply drop from 7.5 ks to $\sim$ 40 ks with a power law index of $\alpha=4.9\pm0.2$ \citep{2013ApJ...766...30G}; (III) a noticeable X-ray bump\citep{2013ApJ...766...30G,Yu2015}, similarly resemble of the fall-back bump seen in GRB 070110 \citep{2017ApJ...849..119C}; (IV) a final shallow decay, to be usually interpreted as standard external
forward shock emission.

\begin{figure*}[htp]
\centering
\includegraphics[origin=c,angle=0,scale=1,width=0.8\textwidth,height=0.50\textheight]{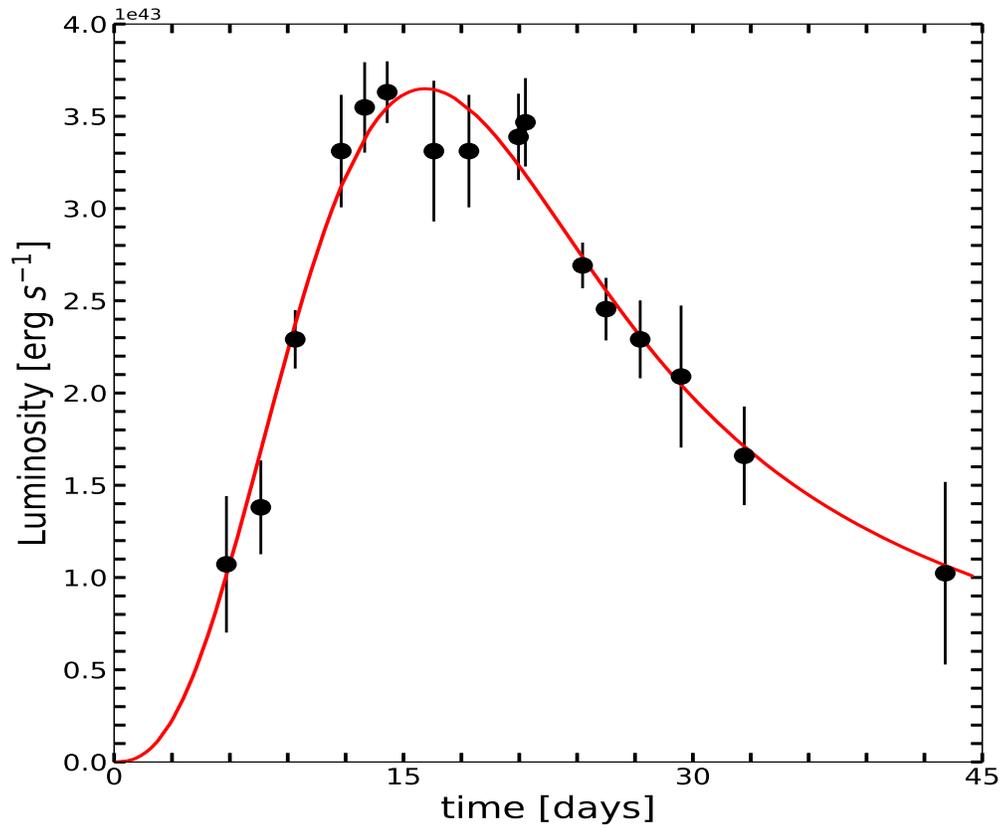}
\caption{ Modeling result for the light curve of SN 2011kl from \cite{2019A&A...624A.143K}. The solid line denotes the BP-powered model.}\label{fig:SN2011kl}
\end{figure*}

By comprehensive analyses of the data from the X-shooter instrument on the Very Large Telescope \citep{Vernet2011} near the peak of the excess emission (29 December 2011),  \cite{2015Natur.523..189G} found that SN~2011kl is spectroscopically associated with ultra-long GRB 111209A.
Their further investigations showed that this supernova with a peak luminostiy $\sim 3.63\times10^{43}$ erg s$^{-1}$ is more than three times more luminous than type Ic SNe. The time of peak brightness of SN~2011kl is $\sim$ 14 days, slightly larger than the average (13.0 days with a standard deviation of 2.7 days; \citealt{2017AdAst2017E...5C}) of other GRB-SNe. At the same time, its spectrum appears rather featureless, resembling those of super-luminous SNe, but extends further down into the rest-frame ultraviolet. Next, \citet{2019A&A...624A.143K} performed the supernova fitting together with considering a filter-dependent host-galaxy component at very late times, and derived the bolometric light curve of SN~2011kl, as shown in Fig.~(\ref{fig:SN2011kl}).

\section{The Model applied to GRB~111209A/SN~2011kl}
\subsection{The collapse of magnetar due to r-mode GW radiation}
Phenomenally, a smooth broken power-law function is usually invoked to fit the  X-ray light curve (e.g., \citealt{liang07,2008ApJ...675..528L}). The function reads
\begin{eqnarray}
	 F &=& F_{0}[(\frac{t}{t_{\rm x,b}})^{\omega\alpha_{\rm x,1}}+ (\frac{t}{t_{\rm x,b}})^{\omega\alpha_{\rm x,2}}]^{-1/\omega},
	\end{eqnarray}
where $\omega=3$, describes the sharpness of the break. We here firstly employ the model-independent function to fit the first two segments of the  X-ray light curve of GRB~111209A, and obtain that $\alpha_{\rm x,1}=0.66\pm0.01$, $\alpha_{\rm x,2}=4.03\pm0.05$ and $t_{\rm x,b} =7800$ s (fixed), with the statistical measure value of R-Squared=0.93.

Physically, to capture the properties of the X-ray emission obtained above, we adopt a magnetar model with an r-mode GW radiation as described below in a simple manner (detailed information refers to \citealt{2016MNRAS.463..489H}). 

A newborn magnetar with spin period $P$ can extract extra energy to power a GRB light curve via its rotational energy
	\begin{equation}
	\Erot = I\Omega^2/2 = 5.68\times 10^{52}\mbox{erg }
	(P/1\mbox{ms})^{-2},
	\end{equation}
	where $\Omega$ $(=2\pi/P)$ and $I$ $(\approx 2.88\times10^{45}\mbox{gcm$^2$ })$
	are the magnetar angular spin frequency and moment of inertia, respectively.
	
	If the magnetar spins down via a combination of electromagnetic dipole (MD) and gravitational wave (GW) radiation, then the evolution of the magnetar rotational energy  $E_{\rm rot}$ can be described as
	\begin{equation}
	\Edotrot = I\Omega\dot{\Omega} = \Edotem+\Edotgw ,\nonumber \\ \label{eq:edotrot}
	\end{equation}
where $\Edotem$ and $\Edotgw$ are the energy loss rate due to MD radiation and GW radiation, respectively.
The MD radiation power is given by (see, e.g. \citealt{2016MNRAS.463..489H})
	\begin{eqnarray}
	\Edotem&=&-\beta I\sin^2\theta\Omega^4 \nonumber\\
	&=& -2.87\times 10^{47}\mbox{erg s$^{-1}$ }\sin^2\theta B_{14}^2
	(P/1\mbox{ms})^{-4}, \label{eq:Edotem}
	\end{eqnarray}
where $B_{14}$ is the surface magnetic field strength, $\theta$ is the angle between the MD moment and rotation axis (see,e.g. \citealt{1968Natur.219..145P,1969Natur.221..454G,2006ApJ...648L..51S}),
	$\beta \equiv B^{2} R^6 /6c^{3} I = 6.40 \times 10^{-14} \mbox{s} B_{14}^2 $, and a magnetar with
mass $M=2.5\, M_{\odot}$ and radius $R=12 \mbox{~km}$ is adopted\footnote{\citealt{2014ApJ...781...26A} argued that for any compact NS with mass of $1 M_{\odot} \leq M \leq 2.5 M_{\odot} $, and radius of $10 km \leq R \leq 15 km$, the moment of inertia ($I$) is uncertain within at most an order of magnitude. Further, \citealt{2016MNRAS.463..489H} also pointed out that the differences in the mass and radius will not change the energy-loss rate due to r-mode GW radiation significantly. Therefore our results are not sensitive to these parameters. }. Then the time-scale over which the magnetar loses energy via MD radiation is
\begin{equation}
\label{eq:tem}
\tau_{\rm em} = \left|\frac{\Erot}{\Edotrot}\right|_{\Omega_0}=1.98\times10^{5} \mbox{s} \sin^{-2}\theta B_{14}^{-2}(P_0/1\mbox{ms})^{2}.  \nonumber\\
\end{equation}

Supposing that the magnetar has an r-mode oscillation with a constant saturation amplitude $\alpha$ \citep{1998PhRvD..58h4020O}, which corresponds usually to a highest value of the braking index $n=7$, the energy loss rate and timescale of the GW radiation are respectively \citep{1998PhRvD..58h4020O,2016MNRAS.463..489H}
\begin{eqnarray}
\Edotgwr  &=& -\gammar I\Omega^8  \label{eq:Edotgw}\\
	&=& -4.86\times 10^{47}\mbox{erg s$^{-1}$ }\alpha_{-2}^2 (P/1\mbox{ms})^{-8}\nonumber, \\
	\tgwr &=& \left|\frac{\Erot}{\Edotgwr}\right|_{\Omega_0}\!\! \nonumber \\
	&=& \frac{1}{2\gammar\Omega_0^6}  \nonumber \\
	&=& 1.17\times 10^{5}\mbox{ s }\alpha_{-2}^{-2}  (P_0/1\mbox{ms})^{6},   \label{eq:tgwr}
\end{eqnarray}
where $\gammar\equiv(96\pi/15^2)(4/3)^6(GMR^4\tilde{J}^2/c^7\tilde{I})\alpha^2$.
Note that $\tilde{J}$ $(=0.0341)$ and $\tilde{I}$ $(=I/MR^2=0.4)$ are derived by using a NS with mass $M=2.5\, M_{\odot}$ and radius $R=12\mbox{ km}$.
	
GWs from a neutron star can be generated in two ways, by a static quadrupolar deformation (`mountain') or by a fluid oscillation
\citep{2016MNRAS.458.1660L}. If GW radiation comes from an r-mode oscillation with constant saturation amplitude in the neutron star, Equation (\ref{eq:edotrot}) becomes
\begin{equation}
I\Omega\dot{\Omega} = - \beta I \Omega^4-\gammar I\Omega^8 .\label{Eq:Omegat}
\label{EqOmega}
\end{equation}\begin{flushleft}
	
\end{flushleft}
Dividing both sides of Equation (\ref{Eq:Omegat}) by $I \Omega$, one have
\begin{equation}
\frac{d\Omega}{dt} = -\beta\Omega^3-\gammar\Omega^{7}. \label{eq:omegadotr}
\end{equation}
When r-mode GW radiation losses dominate the spin-down, i.e., the first term on the right-hand side of Equation (\ref{eq:omegadotr}) can be neglected, one can have a solution of Equation~(\ref{eq:omegadotr}) as (also see \citealt{2016MNRAS.463..489H})
\begin{equation}\label{eq:Omegt}
\Omega(t)=\Omega_{0}\left(1+\frac{3t}{\tau_{\rm gw,r}} \right)^{- \frac{1}{6}},
\end{equation}
where $\tau_{\rm gw,r}$ is the characteristic timescale of the spin-down for a magnetar due to GW radiation. Correspondingly, the electromagnetic luminosity can be obtained
by substituting Equation~(\ref{eq:Omegt}) into Equation~(\ref{eq:Edotem}) \citep{2019ApJ...871..160L}, i.e.,
\begin{eqnarray}
L_{\rm x}&=&L_{\rm x,0}\left(1+\frac{3t}{\tau_{\rm gw,r}}\right)^{-\frac{2}{3}}  \label{eq:Lem(t)}
\end{eqnarray}
with
\begin{eqnarray}\label{equ:Lx}
L_{\rm x,0}=2.87\times 10^{47}\mbox{erg s$^{-1}$} \eta f_{b}^{-1} \sin^2\theta B_{14}^2(P_0/1\mbox{ms})^{-4}  \label{eq:Lem(t1)},
\end{eqnarray}
where $\eta$ is the efficiency in converting the spin-down energy into the X-ray luminosity and $f_b$ is the beaming factor of the GRB outflow \citep{2011MNRAS.413.2031M}. In this paper, we take $\theta=\pi/4$ \citep{dai12,2016ApJ...817..132D}.
	
Eventually, the magnetar would evolve to a state that the centrifugal force is not enough to support the mass \citep{2014PhRvD..89d7302L,2019arXiv190501387L}. In this situation, the magnetar will collapse and form a rotating BH.
The rotational energy of the newborn BH can be extracted by the BZ mechanism, which can be characterized by a simple exponential decay law \citep{2016MNRAS.455.4479N}, i.e.,
\begin{eqnarray}
L_{\rm BZ} \propto e^{-t/\tau_{BZ}},
\label{bzmodel}
\end{eqnarray}
where
\begin{eqnarray}
\tau_{BZ} &=& 200(\frac{B_{eh}}{10^{15}})^{-2} (\frac{M}{2.5M_{\odot}})^{-1} s
\label{bztime}
\end{eqnarray}
is the time-scale for the spinning down procedure and $B_{\rm eh}$ is the magnetic field around the event horizon.
	
Then, to reproduce the initial X-ray afterglow emissions of GRB~111209A, we resort to the following model-dependent functon,
	\begin{eqnarray}\label{Equationtwomodel}
	L_{\rm x}(t)&=& \cases{
		L_{\rm x,0}\left(1+ \frac{3t}{\tau_{\rm gw,r}}\right)^{-\frac{2}{3}} \hspace{1.2cm}      t < t_{\rm col} \cr
		L_{{_{\rm{BZ},0}}}\exp({-\frac{t}{\tau_{_{\rm BZ}}}}) \hspace{1.6cm}  t \geq t_{\rm col} \cr
	},
	\end{eqnarray}
where $t_{\rm col}$ is the time when the magnetar collapses to a BH and is a free parameter in our fitting. The python package of {\tt lmfit}
\footnote{https://lmfit.github.io/lmfit-py/} is used to perform non-linear optimization analysis, in which the {\tt emcee} function is invoked to search the best fit parameters for our model.
The best fitting parameters are read as $L_{\rm x,0}=(5.05\pm0.55)\times10^{49}$erg s$^{-1}$, $t_{\rm col}=7488\pm496$ s, $\tau_{\rm gw,r}=265\pm45$ s, and $\tau_{\rm BZ}=3296\pm20$ s with $\chi^2_{r}=1.48$. One would see that the initial smoothly shallow decay phase with a random superimposition of some larger pulse structures (as seen discussion below), and a few larger pulses would produce larger $\chi^2_{r}$ \citep{2018ApJ...865..153L}. Fig. (\ref{fig:GRB111209A}) shows the best model's compatibility with the data, in which the red solid line represents the magnetar model with an r-mode gravitational-wave emission (equation \ref{eq:Lem(t)}) while the magenta color dotted line shows the component described by equation \ref{bzmodel}.

With the resulting characteristic timescale for r-mode gravitational-waves spindown, $\tau_{\rm gw,r}=265\pm45$ s, we obtain the initial spin period of the magnetar, $P_0=0.36\sim0.78$ ms (this period is much similar to that obtained from GRB~090510 in \citealt{2019ApJ...871..160L}) by adopting typical saturation amplitude $\alpha=0.01\sim0.1$ (please referred to the last section for much more detailed discussion)\citep{2014ApJ...781...26A,2016MNRAS.463..489H,2016ApJ...817..132D}. Assume that the mass of the newborn BH inherits with that of the original magnetar with $2.5M_{\odot}$, we obtain the magnetic field on the event horizon, $B_{\rm eh} \sim 3.5\times 10^{14} G$.

\subsection{Fall-back Accretion and Modeling the Light Curve of SN~2011kl }
When a supramassive magnetar collapses to a BH, a fraction of the envelope materials would fall back and activate the accretion onto the BH at late times, which is widely discussed in the GRB literature \citep{2008MNRAS.388.1729K,2008Sci...321..376K,2013ApJ...767L..36W,2016ApJ...826..141G,2017ApJ...849..119C}. Consequently, the X-ray bump of some GRBs at late times has been considered as fall-back accretion onto
the new-born BH. Some analytical and numerical calculations suggested that the fall-back accretion rate initially increases with time as $\dot{M}_{\rm early} \propto t^{1/2}$ until it reaches a peak value at $t_{\rm p}$ \citep{macfadyen01,zhang08,dai12}. Following \citet{chevalier89}, the late-time fall-back accretion behavior would follows $\dot{M_{\rm late}}\propto t^{-5/3}$. Therefore, we use a smooth broken power-law function of time to describe the evolution of the fall-back accretion rate as \citep{2013ApJ...767L..36W}
	\begin{eqnarray}
	\dot{M} = \dot{M}_{\rm p} \left[ \frac{1}{2}\left(\frac{t-t_0}{t_{\rm p}-t_0} \right)^{-\alpha_{r}s} +  \frac{1}{2}\left(\frac{t-t_0}{t_{\rm p}-t_0} \right)^{-\alpha_{d}s} \right]^{-1/s},\label{eq:dotM}
	\end{eqnarray}
	where $\alpha_{r}=1/2$, $\alpha_{d}=-5/3$, and $s$ describes the sharpness of the peak. We take $s=3$ in this paper.
	
The BZ jet power from a Kerr BH with mass $M_\bullet$ and (or dimensionless mass $m_{\bullet}=M_{\bullet}/M_\odot$) \citep{blandford77} and angular momentum $J_\bullet$ can be given by \citep{lee00,li00,wang02,mckinney05,lei08,lei11,lei13}
\begin{equation}
	L_{\rm BZ}=1.7 \times 10^{50} a_{\bullet}^2 m_{\bullet}^2
	B_{\bullet,15}^2 F(a_{\bullet}) \ {\rm erg \ s^{-1}},
\label{eq_Lmag}
\end{equation}
where $a_\bullet = J_\bullet c/(GM_\bullet^2)$ is the BH spin parameter and $F(a_{\bullet})=[(1+q^2)/q^2][(q+1/q) \arctan q-1]$ with $q= a_{\bullet} /(1+\sqrt{1-a^2_{\bullet}})$. $B_{\bullet}$ is the magnetic field strength around the BH horizon. Since a massive torus of material holds the magnetic field on the event horizon, one can estimate its value by equating the magnetic pressure on the horizon to the ram pressure of the accretion flow at its inner edge \citep{moderski97},
\begin{equation}
\frac{B_{\bullet}^2}{8\pi} = P_{\rm ram} \sim \rho c^2 \sim \frac{\dot{M} c}{4\pi r_{\bullet}^2},
\label{Bmdot}
\end{equation}
where $r_{\bullet}=(1+\sqrt{1-a_\bullet^2})r_{\rm g}$ is the radius of the BH horizon and $r_{\rm g} = G M_\bullet /c^2$. We can then obtain the BZ power as a function of mass accretion rate and BH spin as
\begin{equation}
L_{\rm BZ}=9.3 \times 10^{53} a_\bullet^2 \dot{m}  X(a_\bullet) \ {\rm erg \ s^{-1}}
\label{eq:EB}
\end{equation}
with
\begin{equation}
X(a_\bullet)=F(a_\bullet)/(1+\sqrt{1-a_\bullet^2})^2.
\end{equation}
The observed X-ray luminosity is connected to the BZ power through
\begin{equation}
\eta_{\rm bz} L_{\rm BZ}=f_{\rm b,bz} L_{\rm X,iso},
\label{eq1}
\end{equation}
where $\eta_{\rm bz}$ and $f_{\rm b,bz}$ are efficiency of converting BZ power to X-ray radiation and the jet beaming factor, respectively.
	
The BH may be spun up by accretion and spun-down by the BZ mechanism. For a Kerr BH in the BZ model, the evolution equations are given by \citep{wang02}
\begin{equation}
\frac{dM_\bullet c^2}{dt} = \dot{M} c^2 E_{\rm ms} - \dot{E}_{\rm B}
\label{eq:dMbz}
\end{equation}
and
	\begin{equation}
	\frac{dJ_\bullet}{dt} = \dot{M} L_{\rm ms} - T_{\rm B},
	\label{eq:dJbz}
	\end{equation}
	where $T_{\rm B}$ is the total magnetic torque described as
	\begin{eqnarray}
	T_{\rm B} =3.36 \times 10^{45} a_\bullet^2 q^{-1} m_{\bullet}^3 B_{\bullet,15}^2  F(a_\bullet){\rm \ g \ cm^2 \ s^{-2}}.
	\end{eqnarray}
	
Considering $a_\bullet = J_\bullet c/(GM_\bullet^2)$, by incorporating Equations~(\ref{eq:dMbz}) and (\ref{eq:dJbz}), the evolution of the BH spin can be expressed by
	\begin{eqnarray}
	\frac{da_\bullet}{dt} = && (\dot{M} L_{\rm ms} - T_{\rm B})c/(GM_\bullet ^2) - \nonumber \\
	&& 2 a_\bullet (\dot{M} c^2 E_{\rm ms} - \dot{E}_{\rm B}) /(M_\bullet c^2),
	\label{eq:da}
	\end{eqnarray}
where
\begin{equation}
	E_{\rm ms} = \frac{4\sqrt{ R_{\rm ms} }-3a_{\bullet}}{\sqrt{3} R_{\rm ms}}
\end{equation}
and
\begin{equation}
	L_{\rm ms} = \frac{G M_\bullet}{c} \frac{2 (3 \sqrt{R_{\rm ms}} -2 a_\bullet) }{\sqrt{3} \sqrt{R_{\rm ms}} }
\end{equation}
are the specific energy and angular momentum corresponding to the inner most radius $r_{\rm ms}$ of the disk \citep{1973blho.conf..343N}, respectively.
Here, $R_{\rm ms} = r_{\rm ms}/r_{\rm g}$ is the radius of the marginally stable orbit in terms of $r_{\rm g}$
 and can be described as \citep{bardeen72},
	\begin{eqnarray}
	R_{\rm ms} =  3+Z_2 -\left[(3-Z_1)(3+Z_1+2Z_2)\right]^{1/2}
	\end{eqnarray}
for $0\leq a_{\bullet} \leq 1$, where $Z_1 \equiv 1+(1-a_{\bullet}^2)^{1/3} [(1+a_{\bullet})^{1/3}+(1-a_{\bullet})^{1/3}]$ and $Z_2\equiv (3a_{\bullet}^2+Z_1^2)^{1/2}$.
	
For Equation (\ref{eq:EB}), one can calculate the peak accretion rate
\begin{equation}
\dot{M}_{\rm p} \simeq 1.1 \times 10^{-9} L_{\rm X,iso,45} a_\bullet^{-2} X^{-1}(a_\bullet)  \eta_{\rm bz,-2}^{-1} f_{\rm b,bz,-2} M_{\sun}~{\rm s}^{-1},
\end{equation}
where $\eta_{\rm bz,-2} =\eta_{\rm bz}/10^{-2}$ and $f_{\rm b,bz,-2}=f_{\rm b,bz}/10^{-2}$.
From Equation(\ref{eq:dotM}), we can estimate the total fallback/accreted mass \citep{2013ApJ...767L..36W}
\begin{eqnarray}
M_{\rm fb}&& \simeq \int_{t_0}^{t_p} \dot{M}dt \sim 2\dot{M}_{\rm p} (t_{\rm p} -t_0)/3  \nonumber \\
	&& \simeq  4.6 \times 10^{-7} L_{\rm X,iso,45} a_\bullet^{-2} X^{-1}(a_\bullet)  \eta_{\rm,bz,-2}^{-1} f_{\rm b,bz,-2} M_{\sun}. \nonumber
\end{eqnarray}
In the following,
we fit the fallback accretion model and extra power-law (PL) function to the late-time small X-ray bump in GRB~111209A by the python package {\tt lmfit} as above, where PL function can be responsible for the standard external forward shock component, and the best parameters are $t_{\rm 0}=27000$~s (fixed), $t_{\rm p}=57624\pm5490$ s, $L_{\rm x,iso}=(1.95\pm0.37)\times10^{45}$ erg$s^{-1}$, and the PL index $\Gamma=-1.06\pm0.03$ with $\chi_{r}^2=0.989$. The best models are also shown in Fig.~(\ref{fig}).

The initial setup for a newborn BH can be obtained by assuming that the newborn BH inherits the mass and angular momentum from the supramassive magnetar. With the above parameters for a magnetar, we get an initial BH mass $m_\bullet=2.5$ with a initial spin $a_\bullet \sim 0.02$ by using $J_\bullet=2 \pi I/P_{\bullet}$, where $P_{\bullet}$ corresponding to the late time ($t_{0} \sim 27000 s$).  According to fallback accretion model, we obtained that the the peak accretion rate and total fall-back mass are $\dot{M}_{\rm p} = 6.43 \times 10^{-5}M_\sun$ and $M_{\rm fb}=1.31 M_\sun$ by considering the radiation efficiency $\eta_{\rm bz}=0.01$ and beaming factor $f_{\rm b,bz}=0.02$ (\citealt{2013ApJ...767L..36W,2016ApJ...826..141G}), respectively. The outermost radius of the fallback material could be essentially estimated as $r_{\rm fb}=5.9 \times 10^{11}$ cm, suggesting that the progenitor of GRB~111209A may be a Wolf-Rayet stars, the same as proposed by previous authors (\citealt{2015Natur.523..189G,2016ApJ...826..141G}).

If the X-ray bump is due to the contribution from the BZ jet according to the fallback BH disk. Then when some envelope materials fall back and activate the accretion onto the BH,
a centrifugally driven baryon-rich wide wind/outflow from the revived accretion disk through the BP mechanism would deposit energy on the SN ejecta, which produces the high luminosity of SN~2011kl as argured in \cite{2016ApJ...826..141G}. The BP outflow luminosity could be estimated as \citep{AN99,2016ApJ...826..141G,ma2018}
\begin{eqnarray}
L_{\rm BP}=\frac{(B_{\rm ms}^{\rm P})^2r_{ms}^4\Omega_{ms}^{2}}{32c},
\end{eqnarray}
where $B_{\rm ms}^{\rm P}$ and $\Omega_{ms}$ are the poloidal disk field and the Keplerian angular velocity at inner stable circular orbit radius ($r_{ms}$), respectively.
The angular velocity $\Omega_{\rm ms}$ is given by
\begin{eqnarray}
\Omega_{\rm ms}=\frac{1}{M_\bullet/c^3(\chi^{3}_{\rm ms}+a_{\bullet})}
\end{eqnarray}
where $\chi_{\rm ms}\equiv \sqrt{r_{\rm ms}/{M_\bullet}}$. The the poloidal disk field, $B_{\rm ms}^{\rm P}$, is given by the following formula \citep{blandford82},
\begin{eqnarray}
B_{\rm ms}^{\rm P}=B_{\bullet}(r_{ms}/r_H)^{-5/4},
\end{eqnarray}
where $r_{H}=M_\bullet(1+\sqrt{1-a_{\bullet}^2})$ is the horizon radius of the BH.
	
When the initial total thermal energy of a SN could be neglected, the photospheric luminosity of the SN powered by a variety of energy sources could be expressed as (detailed information refers to \citealt{arnett82,wang15a,wang15b})
\begin{eqnarray}
L(t)&=&\frac{2}{\tau_{d}}e^{-\left(\frac{t^{2}}{\tau_{d}^{2}}+\frac{2R_{0}t}{v\tau_{d}^{2}}\right)}~
\int_0^t e^{\left(\frac{t'^{2}}{\tau_{d}^{2}}+\frac{2R_{0}t'}{v\tau_{d}^{2}}\right)}\left(\frac{R_{0}}{v\tau_{d}}+\frac{t'}{\tau_{d}}\right)   \nonumber\\
	&&\times L_{\rm inj}(t')dt',
\label{equ:LumSN}
\end{eqnarray}
where $R_{0}$ and $L_{\rm inj}$ are the initial radius of the progenitor and the generalized power source, respectively. Here we take $L_{\rm inj}=L_{\rm BP}$.  $\tau_{d}=(2\kappa M_{\rm ej}/{\beta vc})^{1/2}$ is the effective light curve timescale, where $\kappa=0.07\rm  cm^{2}~g^{-1}$, $M_{\rm ej}$ and $v$ are the Thomson electron scattering opacity, the ejecta mass, and the expansion velocity of the ejecta, respectively. $\beta \simeq 13.8$ is a constant that accounts for the density distribution of the ejecta.

Then, we directly fit Equation (\ref{equ:LumSN}) to the bolometric light curve of SN~2011kl through the same method as above.
Being suffer severe degeneracy between the two parameters, $M_{\rm ej}$ and $v$, we fix a value of $v\sim20000$ km/s obtained by (\citealt{2015Natur.523..189G,2017ApJ...850..148W,2019A&A...624A.143K}) in their search of best fit parameters, and then yield that, $t_{\rm p}=993786\pm342731$ s, $L_{\rm x}=(4.745\pm1.656)\times10^{43}$ erg $\rm s^{-1}$, and $M_{\rm ej}=3.22\pm 1.47 M_{\odot}$. Shown in Fig.~(\ref{fig:SN2011kl}) is the best modeling result for the light curve of SN~2011kl.

\section{Summary and discussions}

Based on the observed features exhibiting in the X-ray light curve of GRB~111209A, coupled with the temporal and spectral features observed in the unusual super-luminous SN~2011kl, we give the physical picture for GRB~111209A/SN~2011kl as follows: the central engine of ultra-long GRB~111209A is initially a supramassive magnetar. As its angular momentum is lost duing to gravitational radiation, the magnetar spins down, which powers the initial shallow X-ray emission. Until its centrifugal force supply is insufficient to support the mass, the magnetar collapses to form a BH, then the electromagnetic extraction of spin energy from the newborn BH results in the steeply declining X-ray flux. Next, a fraction of the envelope materials falls back and activates the accretion onto the newborn BH, which produces the X-ray rebrightening bump at late times. At the same time, the centrifugally driven baryon-rich quasi-isotropic wind from the revived accretion disk via BP mechanism deposits its kinetic energy on the SN ejecta, which powers super-luminous SN~2011kl. 

\begin{figure*}[htp]
\centering
\includegraphics[origin=c,angle=0,scale=1,width=0.8\textwidth,height=0.50\textheight]{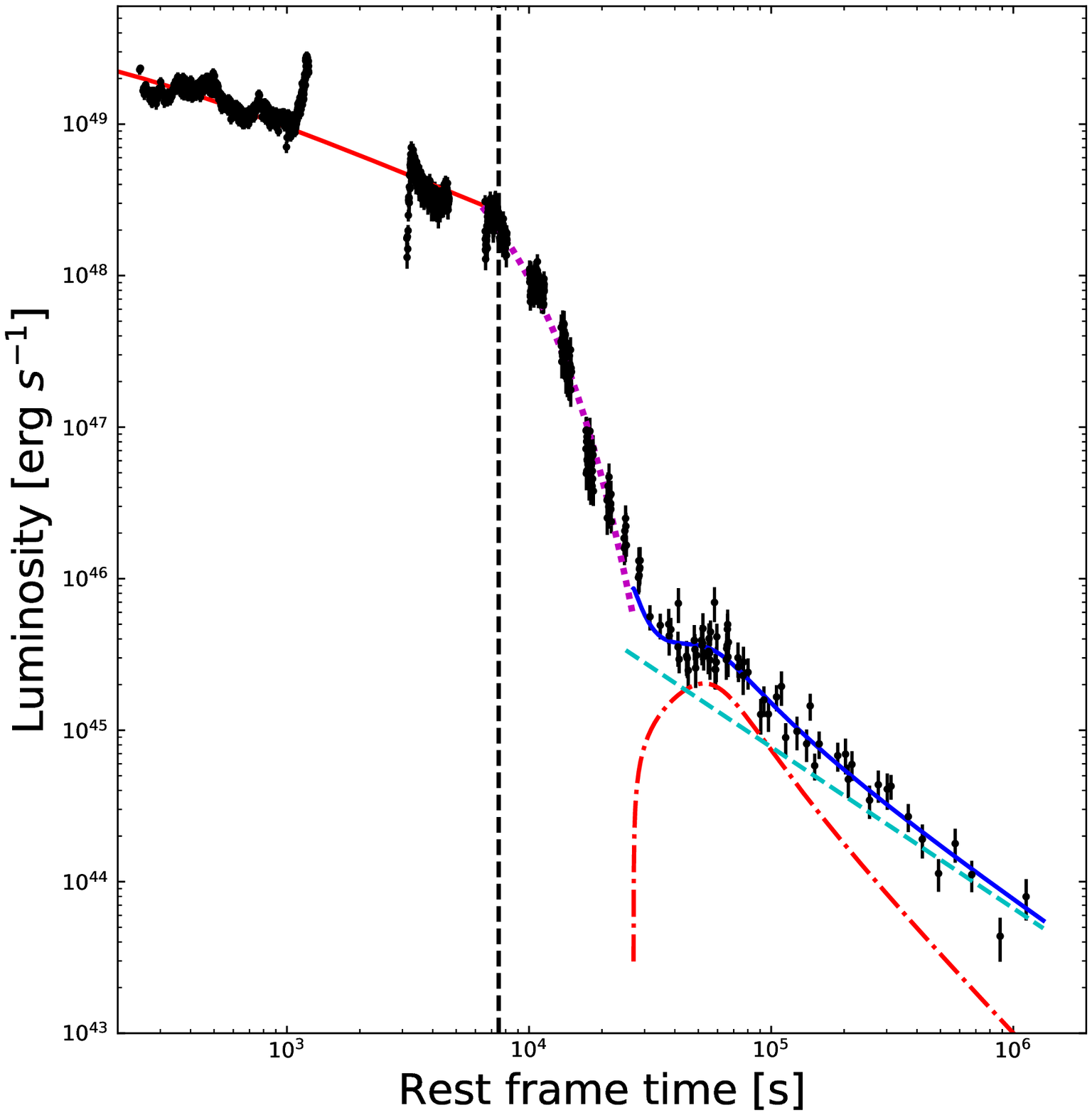}
\caption{\label{fig} Demonstrations of the compatibility of the best-fit our models with the X-Ray light curve of GRB 111209A in the rest-frame. Red solid line is the magnetar model which r-mode gravitational-wave emission dominant the spin down, the magenta dot-dashed line is the theoretical exponential newborn BH spin down. The red dash-doted line and cyan dashed line are the fallback accretion model and the standard external forward shock component, respectively, while the blue solid line denotes the sum of the two emission components. The vertical dash line marks the location of the collapse time $t_{\rm col}$.}\label{fig:GRB111209A}
\end{figure*}

Benefited from the high quality of XMM-Newton data, \cite{2013ApJ...779...66S} found that there exists a shallow decay phase with decay slope of $\sim0.18$ following the sharply decay phase. Thus, the preference of an addition of a second harder component over the standard power-law soft spectrum is proposed to accommodate the data in the shallow decay phase.
The observational feature could be naturally explained in the context of the physical picture mentioned above. One could find from Fig. (\ref{fig:GRB111209A}) that the combination of the two components, external forward shock and fallback accretion component, contributes to the shallow temporally decay phase ($\sim(3-7)\times 10^4$ s), i.e., the additional hard power-law component comes from the fallback accretion on the newborn BH, while the soft power-law component from the standard external forward shock.

To test whether the initial shallow  X-ray emission favors a fireball external shock model, we extracted the spectral index of $\beta_{x,1}=0.17 \pm 0.01$ using Swift/XRT data in the time interval of $[T_{0}+253, T_{0}+1073]$ s from the Swift online repository (\citealt{2009MNRAS.397.1177E}). According to the synchrotron closure relationship between the temporal and spectral indices (e.g.,\citealt{2007ApJ...662.1093W,2013ApJ...779...66S}), we obtain a much shallower expected temporal index during this phase, $\alpha_{x,exp1} = 3\beta_{x,1}/2 = 0.26 \pm 0.01$ or $\alpha_{x,exp1} =(3\beta_{x,1}-1)/2 = {-0.16} \pm 0.01$ for $\nu_{m} < \nu < \nu_{c}$ or $\nu > \nu_{c}$ , respectively, comparing to the observed temporal index ($\alpha_{x,1} = 0.66\pm0.01$), which suggests that the magnetar central engine mention above is favored over the fireball external shock model.	

Alternatively, there are other interpretations for a sharply decay phase observed in X-ray light curve of GRBs, such as reverse shock emission or high latitude emission. Here we perform a test in a simple situation. We firstly extracted the spectral index of $\beta_{x,2}=0.71^{+0.078}_{-0.077}$ in the sharply decay phase of GRB~111209A from the Swift online repository (\citealt{2009MNRAS.397.1177E}). According to a generic external shock model, electrons in the reverse shock region can scatter the synchrotron photons (synchrotron self-inverse Compton emission, SSC) to the early X-ray (e.g., \citealt{2007ApJ...655..391K,2017arXiv171008514F}). With the derived electron spectral index $p=2.42 ^{+0.16}_{-0.15}$ , one could obtain the predicted temporal decay index of $\alpha_{x,exp2}=(3p+1)/3=2.75^{+0.16}_{-0.15}$ for $\nu_{m}^{IC}< \nu < \nu_{c}^{IC}$ in the case of thin shell, whereas in the case of thick shell, the expected temporal decay index would be $\alpha_{x,exp2}=(5p+1)/5=2.62^{+0.16}_{-0.15}$. Clearly, the observation does not support the reverse shock interpretation. Even for the situation of high latitude radiation dominates \citep{2000ApJ...541L...9K,2006ApJ...642..354Z,2007ApJ...655..973K}, one also would obtain a shallower temporal decay index of $\alpha_{x,exp2}=2+\beta_{x,2}=2.71^{+0.078}_{-0.077}$ comparing to the observed temporal index of $\alpha_{x,2}= 4.03\pm0.05$. However, we also note that the value of $\alpha_{x,2}$  is very sensitive to the assumed zero time point $t_{0}$, which is directly constrained from the data \citep{2006ApJ...642..354Z,2007ApJ...655..973K}. Therefore, we could not rule out the curvature effect interpretation.
	
\begin{figure*}[htp]
\centering
\includegraphics[origin=c,angle=0,scale=1,width=0.8\textwidth,height=0.50\textheight]{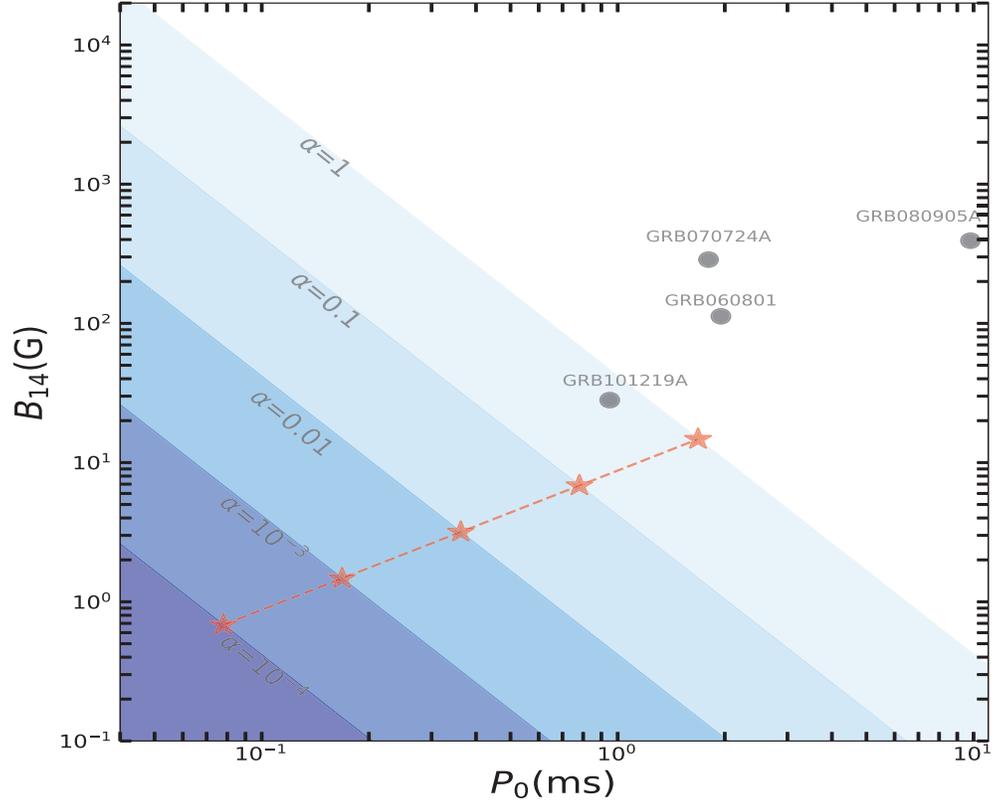}
\caption{Constraints on magnetar's surface magnetic field strength $B$ based on Equation (\ref{eq:B_P0}) as a function of it's initial spin period $P_{\rm 0}$ based on the observation of GRB~111209A. Stars present the different initial spin period $P_{\rm 0}$ corresponding to different r-mode amplitude ($\alpha=10^{-4},10^{-3},10^{-2},0.1,1,$ from left to right, as marked in the figure) derived from Equation (\ref{eq:tgwr}) with the timescale $\tau_{\rm gw,r}=265$ s. Circles are $P_{\rm 0}$ and $B$ for four short GRBs from \citealt{2014PhRvD..89d7302L,2013MNRAS.430.1061R}. }
\label{fig:B_p0}
\end{figure*}

Usually, a newborn magnetar spins down through a combination of MD and GW radiation (e.g., \citealt{2018PhRvD..98d3011S,2019ApJ...886....5S}). If a magnetar is a strong enough emitter of GWs, the GW energy loss will cause the magnetar spin frequency to decrease at a faster rate than by pure MD radiation at early times, thus the shape of the light curve of GRB powered by the rotational energy of the magnetar will be significantly altered from that expected from the model which only considers MD radiation \citep{2016MNRAS.463..489H}. Therefore, GW signal could be indirectly identified from the evolutional features observed in the early-time  X-ray afterglow of GRBs \citep{2013PhRvD..88f7304F,2016MNRAS.463..489H,2019ApJ...871..160L}. At same time, one could place a constraint on the magnetar's surface magnetic field strength. Our analysis above shows that the model of an r-mode GW radiation from a spinning down magnetar agrees with the early-time smooth X-ray afterglow observation of GRB~111209A, which indicates that the GW radiation energy-loss rate should be larger than MD radiation energy-loss before the megnetar collapses to form a BH, as pointed out by \cite{2016MNRAS.463..489H}. According to equations (\ref{eq:Edotem}) and (\ref{eq:Edotgw}), we get
\begin{eqnarray}
\frac{\Edotgwr}{\Edotem} &=& 1.69\sin^{-2}\theta \alpha_{-2}^2 B_{14}^{-2}(P_{0}/1\mbox{ms})^{-4}(1+\frac{3t_{\rm col}}{\tau_{\rm gw,r}})^{-\frac{2}{3}}>1\nonumber. \\
\label{eq:B_P0}
\end{eqnarray}
Based on equation (\ref{eq:B_P0}) with the information from observation of GRB~111209A, $t_{\rm col}$ and $\tau_{\rm gw,r}$,  Fig. \ref{fig:B_p0} shows the relationship between magnetic field strength $B$ and initial spin period $P_{0}$ for given $\theta$ and $\alpha$. By taking a typical value of $\theta=\pi/2$ \citep{2019arXiv191014336L}, we could limit the physical parameters of the magnetar powering GRB~111209A to a narrow range, i.e., 0.36 ms $\leq P_{0} \leq$ 0.78 ms and 3.1 G $\leq B_{14} \leq$ 6.8 G, if a typical range of  $0.01 \leq \alpha \leq 0.1$  is adopted (detail information refered to \citealt{1998PhRvD..58h4020O,Yu09,Alford2012,2016ApJ...817..132D}). Interestingly, with the characteristic timescale $\tau_{\rm BZ}=3296 s$ (see section 3.1), we obtained the magnetic field on the event horizon of the newborn BH, $B_{14}\simeq 3.5$ G, which is consistent with that obtained from the magnetar, suggesting that the newborn BH should also inherit the magnetic field from the supramassive magnetar in the initial setup of the newborn BH, in addition to inherit the mass and angular momentum from the supramassive magnetar (\citealt{2017ApJ...849..119C}).

The values of $B$ and $P_{0}$ of magnetars that powers short GRBs, could also be derived by fitting a simple spin-down model to the X-ray plateau observed in the afterglows of some short GRBs (see \citealt{2013MNRAS.430.1061R}). For a comparison, we also plot the data onto Fig. \ref{fig:B_p0}, and find that the $B$ of GRB~111209A is smaller than those of four short GRBs with redshift measurements. Does it suggest that long GRBs tend to have weaker magnetic field than short GRBs? Theoretically, although a merger remnant in short GRBs is typical expected to have stronger magnetic field which is amplified due to the magneto-rotational instability or dynamo as the objects merge, it needs to be furtherly investigated in a larger sample in the future.

It is worth noting that there are two or three bright X-ray flares overlapped on the shallow decay phase of GRB~111209A, which usually is discovered in the afterglow phase of GRBs and is known as a canonical component in GRB X-ray afterglow. Those flares would mess the underlying smooth component and produce a large $\chi_r^2$ during our fitting routines (also seen in \citealt{2018ApJ...865..153L}). This smooth component is commonly linked to the magnetar spin-down radiation with different braking indexes (e.g., \citealt{2016MNRAS.463..489H,Lasky2017,LV2017,2019ApJ...871..160L,2019ApJ...872..114S,2019ApJ...871...54L,2019ApJ...878...62X}). \textbf{ A variety of efforts have been conducted to explain the physical origin of the X-ray flares (e.g., \citealt{Perna2006,dai12,Dall'Osso2017,Metzger2018}).  More recently, with a modified version of the magnetar propeller model with fallback accretion, \cite{Gibson2018} provided a good fits to a sample of long GRBs with bright flares in their X-ray light curves. Magnetohy-drodynamic simulations (\citealt{Toropina2008}) show that if the magnetospheric radius $r_{m}$ is larger than the co-rotation radius $r_{c}$ and smaller than the light cylinder radius  $r_{\rm cl}$, the system enters a propeller regime, thus bright X-ray flares could be powered by the delayed onset of the propeller regime (also see \citealt{1975A&A....39..185I,2013ApJ...775...67B,2014MNRAS.438..240G,Gibson2017}), in which in-falling material is accelerated to super-Keplerian velocities via magneto-centrifugal slinging and is ejected from the system. When the fallback accretion rate is so small that $r_{m}$ exceeds  $r_{\rm cl}$, both the fallback accretion and the propeller effect stop and the torque exerted on the magnetar by the accretion disk disappears (\citealt{dai12}). \cite{Metzger2018} also argued that accretion should cause the magnetar to grow in mass, but how the star actually accepts the matter being fed from the disk depends on some uncertain factors, such as whether the polar accretion column is able to cool through neutrinos and settle on the magnetar surface (\citealt{Piro2011}). Another uncertainty is the efficiency with which matter accretes in the propeller regime. In this paper, we neglect possible accretion on to the magnetar and assume a constant central engine in mass (also see \citealt{2016MNRAS.463..489H}). Centainly, which would make the magnetar take less time to collapse. So we give an caution to readers who wants to use the collapse time $t_{\rm col}$ we obtain above.  }

\acknowledgments

We acknowledge the use of the public data from the Swift and the UK Swift Science Data Center. We appreciate Bing Zhang, He Gao and Liu Tong for their helpful comments and suggestions. This work is supported by the National Natural Science Foundation of China (grant Nos. U1938106, 11533003, 11573034, 11773007 and 11673006), the Guangxi Science Foundation (grant Nos. 2018GXNSFDA281033, 2018GXNSFFA281010, 2016GXNSFDA380027 and 2016GXNSFFA380006), and the Special Funding for Guangxi Distinguished Professors (Bagui Yingcai \& Bagui Xuezhe, grant No. AD17129006).

%



\software{Xspec \citep{1999ascl.soft10005A}, Scipy \citep{scipy}, Matplotlib \citep{Hunter:2007}}





\clearpage
	
\end{document}